# Research on Comprehensive Classroom Evaluation System Based on Multiple AI Models

Xie Cong[1], Yang Li[1], Wang Daben[2], Xiao Jing[2]

[1]Dalian University of Technology [2]Hebei University of Technology

## Abstract

The promotion of the national education digitalization strategy has facilitated the development of teaching quality evaluation towards all-round, process-oriented, precise, and intelligent directions, inspiring explorations into new methods and technologies for educational quality assurance. Classroom teaching evaluation methods dominated by teaching supervision and student teaching evaluation suffer from issues such as low efficiency, strong subjectivity, and limited evaluation dimensions. How to further advance intelligent and objective evaluation remains a topic to be explored. This paper, based on image recognition technology, speech recognition technology, and AI large language models, develops a comprehensive evaluation system that automatically generates evaluation reports and optimization suggestions from two dimensions: teacher teaching ability and classroom teaching effectiveness. This study establishes a closed-loop classroom evaluation model that comprehensively evaluates student and teaching conditions based on multi-dimensional data throughout the classroom teaching process, and further analyzes the data to guide teaching improvement. It meets the requirements of all-round and process-oriented classroom evaluation in the era of digital education, effectively solves the main problems of manual evaluation methods, and provides data collection and analysis methods as well as technologies for relevant research on educational teaching evaluation.

## Keywords

## 1. Research Background

The classroom occupies a core and fundamental position in curriculum teaching. Classroom teaching quality evaluation, as the "command baton" of education and teaching, affects multiple key areas such as student development, teacher growth, and educational strategic planning. In the current domestic educational context, classroom evaluation mainly relies on subjective evaluation methods such as teaching supervisors' class observations, investigations, and student teaching evaluations to ensure teaching quality and promote teaching development. With social progress and development, this subjective evaluation method also needs further reform due to its own limitations, such as: ① Teaching supervisors can only complete the evaluation of the current course within a unit class period, resulting in high cost and low efficiency of classroom evaluation; ② The knowledge reserve and social experience of teaching supervisors have a certain impact on the objectivity of evaluation results; ③ There is a certain error between the contact evaluation results of teaching supervisors and the real situation; ④ Random sampling evaluation lacks the collection of data throughout the teaching process, resulting in limited understanding, evaluation, and suggestions for teachers.

The educational evaluation principle advocated by the state of "making full use of information technology, strengthening process evaluation, exploring value-added evaluation, and improving comprehensive evaluation" determines the development direction of intelligence in the education system. There is an urgent need for new methods that can not only meet the digital education evaluation principles but also make up for the shortcomings of traditional evaluation methods. This study, based on artificial intelligence technology, establishes a digitally empowered comprehensive classroom evaluation system to achieve automatic data collection, evaluation, and analysis of students' performance and teachers' teaching behaviors throughout the classroom teaching process, solve the shortcomings of the current subjective evaluation methods, and meet the needs of process evaluation, multi-dimensional comprehensive evaluation, and exploration of potential value-added in digital education evaluation.

## 2. Current Situation of Classroom Evaluation Development

### 2.1 Related Research on Classroom Evaluation Systems

Classroom evaluation is an evaluation activity carried out for teaching behaviors based on certain evaluation indicators. By recording and analyzing teaching behavior data, it aims to understand the basic situation of the evaluation object and provide optimization suggestions, with the ultimate goals of ensuring and improving educational quality, promoting the comprehensive growth of teachers and students, providing decision-making basis, and promoting curriculum reform. Due to different cultural backgrounds, educational role orientations, and teaching purposes of different schools, many scholars have studied the focus of relevant contents in the classroom evaluation system from different perspectives: in local application-oriented universities, the effectiveness of teaching behaviors and the cultivation of practical abilities should be fully concerned in classroom evaluation [1]; in the evaluation of teaching activities, teacher behaviors and student feedback should be included in the evaluation scope [2], so as to reflect the interaction and support between teaching and learning; when evaluating teaching behaviors, the knowledge system and communication methods become important evaluation indicators, jointly determining the teaching effect [3]. These studies propose the concept of differentiated classroom evaluation in different teaching scenarios, but classroom teaching data is the core supporting the operation of the evaluation system, and all evaluation analyses should be based on classroom teaching activities and processes. For example, comparing and analyzing multiple aspects such as teachers' professional skills, emotional attitudes, and students' experience feedback in classroom teaching data, and having important significance for the improvement of classroom teaching quality according to specific indicators including the integration of curriculum ideology and politics in classroom teaching, the enrichment and compactness of teaching content, the suitability of teaching methods and means, teaching style and attitude, the strengthening of teaching effects, and the evaluation and self-evaluation of learning effects [4-5].

## 2.2 Research Status of Classroom Evaluation Informatization

At present, in the process of teachers' immediate self-evaluation and analysis of teaching effects during teaching, they mainly judge students' interest in the current classroom content or learning difficulty by observing students' behaviors such as looking up, bowing their heads, gazing, and responding to interactions in class, so as to make targeted classroom teaching design adjustments. However, due to the limited energy of teachers, they can only conduct behavior evaluations on part of the students randomly in part of the class period. This not only fails to fully understand the overall learning situation of students, but also the too frequent student behavior evaluation will affect the teaching progress. How to automatically and objectively evaluate the classroom teaching effect and students' learning interest based on students' in-class performance and behavior data is worthy of in-depth exploration and research.

Informatization technology provides data storage and data analysis tools for educational evaluation, making it possible to evolve random sampling evaluation to group analysis, so as to more accurately evaluate the overall status. The relevant research on the teaching evaluation system analyzes the key concepts of contemporary classroom evaluation from different angles [1-3], establishes a dual-object evaluation idea of teacher teaching and student performance, so as to comprehensively understand the classroom teaching process; the evaluation indicators of teachers' teaching should include both knowledge system evaluation and communication form evaluation to ensure teaching quality and effectiveness. Taking students as the evaluation object, computer vision technology can automatically judge students' head movements or facial expressions, analyze students' classroom engagement and learning difficulty, and evaluate teaching effects through students' behavior feedback [6-7]. Taking teachers as the evaluation object, most current colleges and universities mainly adopt the manual supervision evaluation method, hiring retired teachers or school leaders to conduct evaluations through random class listening. A few schools try to use classroom cameras or monitoring equipment to intercept videos or pictures as the basis for teaching evaluation [3]. The outstanding contradiction of the above two evaluation methods for teachers' abilities lies in the insufficient and comprehensive understanding of teaching activities and evaluation objects, which is easy to generalize and take out of context. The key to the problem is the lack of tools for automatic teaching activity data collection, analysis, and induction. At present, the development of computer speech models and large language model related technologies has enabled computers to have audio recognition and language analysis capabilities, and computers can automatically complete the conversion of speech data into text data and the analysis and answering of text content

[8-10]. Therefore, using information technology to complete the automatic data collection, analysis, and induction of the whole teaching process can more comprehensively and truly understand the teaching process, objectively, batchly, and real-timely analyze and evaluate the behaviors of the evaluation objects, and significantly alleviate the outstanding contradictions in classroom teaching evaluation.

**2.3 Research Objectives and Significance**

The iterative development of artificial intelligence technology has further highlighted the urgent need to promote the reform of information-based teaching evaluation. The current information-based application of teaching evaluation is mainly concentrated on the evaluation based on students' performance, while the research on the digitization and informatization of teachers' teaching evaluation is rare, leading to problems such as the lack of understanding of the teaching process, incomplete classroom evaluation, and the inability of evaluation results to guide teaching improvement in a timely manner. Therefore, this paper will use AI empowerment to solve the problems existing in the current teaching process and classroom evaluation.

This paper intends to use computer image recognition technology to establish an information-based evaluation module based on students' classroom behavior performance; use speech recognition technology and AI large language models to establish an information-based evaluation module for teachers' teaching data collection and teaching ability evaluation; through students' classroom behavior data and teachers' teaching data, use large language models to automatically carry out data mapping analysis and generate optimization suggestion reports, and establish an information-based classroom suggestion module. Finally, use the classroom comprehensive evaluation system integrating the three modules of student classroom behavior evaluation, teacher teaching data collection and ability information-based evaluation, and intelligent optimization suggestion report to complete large-scale data collection and analysis, and provide data and decision support for evaluation reform.

# 3. Analysis of Classroom Teaching Process

The classroom teaching is essentially a two-way information communication behavior between teachers and students. Teachers transmit inherent knowledge information to students in the form of speech based on their own knowledge background and expression habits. Students understand and remember the received knowledge information, and teachers maintain or adjust the information transmission strategy according to students' behavioral performance, as shown in Figure 1 Classroom Teaching Flow Chart. The above process goes through several cycles to form the teaching activities of the whole class.

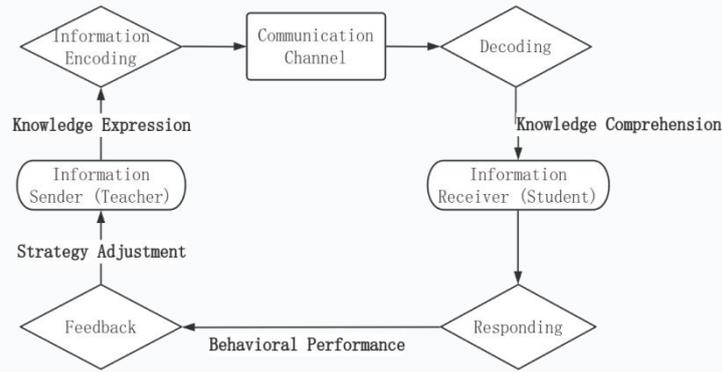

Fig. 1 Flowchart of Classroom Teaching Process.

Whether the information communication is two-way depends on whether the information receiver gives corresponding behavioral feedback to the information sender. It can be seen from Figure 1 that the classroom teaching process can be divided into two main parts: information transmission and behavior feedback:

Information transmission link. Teachers transmit inherent knowledge information to students in the form of speech based on their own understanding depth and expression habits of knowledge information. Students understand and memorize the teacher's expression content based on their own knowledge background. Teachers' own complete knowledge system is the foundation of high-quality teaching. Adopting appropriate expression methods for students' knowledge background to accurately and efficiently transmit knowledge information to students is the realization carrier of high-quality teaching.

Behavior feedback link. Students show different body or facial movements according to their own understanding results of the information. Students' looking up/bowing their heads can prompt the degree and initiative of the student's participation in teaching activities. Through continuous action recognition during the whole class, the adaptability and learning interest of the batch of students to the teaching content and expression methods can be analyzed and judged.

## 4. System Design and Technical Implementation

### 4.1 System Design

The ideal structure of teaching evaluation is a closed-loop process based on a comprehensive understanding of the evaluation object, analyzing the current trends and characteristics, and providing guiding opinions for subsequent practice optimization. Classroom teaching is a closed-loop process in which teachers transmit information to students, and students feedback their learning status through action behaviors. This paper will rely on a variety of artificial intelligence technologies to automatically realize the behavior evaluation of both teachers and students, and establish a closed-loop classroom evaluation system through data mapping analysis to guide education and teaching.

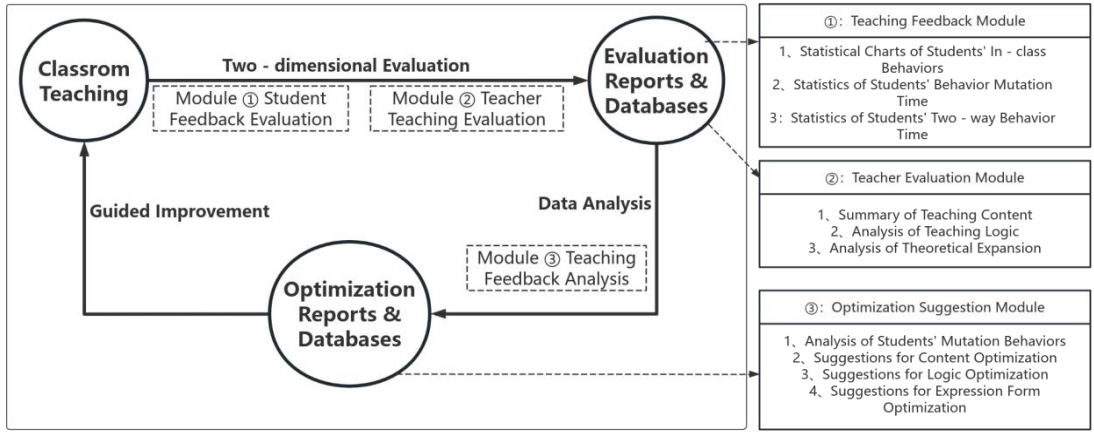

Fig. 2 System framework design diagram.

As shown in Figure 2, the system is mainly composed of three modules: Module 1, the student performance feedback evaluation module, which identifies and counts the students' behavior actions in the classroom, including bowing their heads and looking up, and evaluates the classroom teaching quality from the perspective of whether students participate in classroom learning; Module 2, the teacher teaching process evaluation module, which analyzes the teacher's language content in the classroom, and analyzes the teacher's own teaching ability and evaluates the classroom teaching quality from the perspectives of classroom ideology and politics integration, content logic consistency, theoretical and practical combination degree, subject characteristics, etc.; Module 3, the teaching feedback analysis module, which corresponds the data of Module 1 and Module 2 through the time axis, and analyzes the specific highlights and areas for improvement of the teacher's language content.

## 4.2 Technical Route

In view of the data confidentiality requirements of Chinese universities, all artificial intelligence technologies in this paper adopt a local deployment scheme. The classroom monitoring system is used to obtain classroom teaching videos and synchronous audio as raw data, and the computer is used to realize automatic teaching evaluation. The overall technical route diagram of the system is shown in Figure 3.

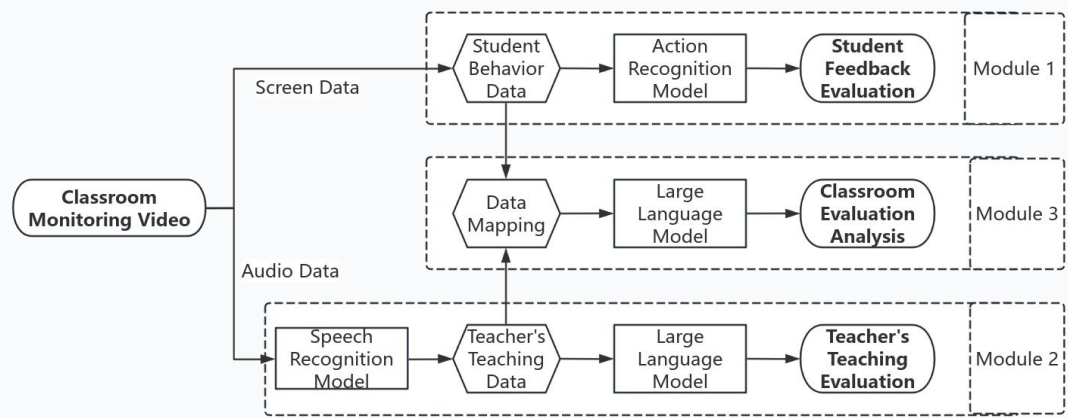

Fig. 3 System flowchart.

### 4.2.1 Module 1

Module 1 establishes an action recognition model for image data based on computer vision technology and deep learning technology, continuously recognizes and records the actions of characters in classroom videos, and finally generates evaluation charts. The accuracy and calculation efficiency of the action recognition model are related to the accuracy and practicability of the evaluation system's evaluation results. The ability of the model is determined by the algorithm, the quality of the training data, and the quantity of the training data together. This paper selects the relatively mature YOLOv8 as the target detection algorithm for establishing the action recognition model. In terms of training data collection, 300 monitoring screenshots of 40-person classrooms, 200 monitoring screenshots of 60-person classrooms, and 100 monitoring screenshots of 120-person classrooms were intercepted, totaling 600 screenshots. In terms of training data annotation, the behavior of each student in 600 screenshots was manually annotated, and finally about 13,500 students looking up and 13,000 students bowing their heads were annotated. Using 600 pictures containing 26,500 students, the YOLOv8 algorithm was used for training and modeling, and the model training results are shown in Figure 4.

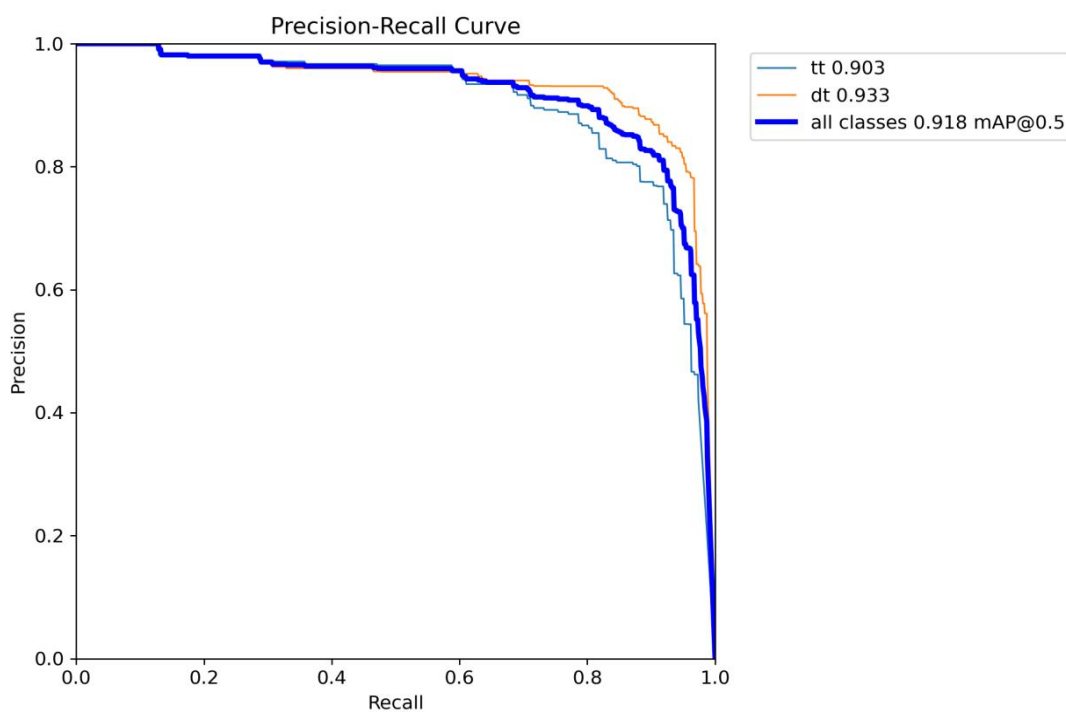

Fig.4 Training result of the action recognition model.

Figure 4 shows that precision represents the proportion of correct recognition results in the model recognition results, and recall represents the proportion of accurately recognized results in all correct results. The mean average precision (mAP) comprehensively considers the precision and recall and is an indicator that can more comprehensively evaluate the model's detection ability. Its value range is 0-1, and the closer it is to 1, the better the model's detection performance for various targets. As shown in Figure 4, the mAP value of the image recognition model after training is

0.918, indicating that the model has excellent recognition ability and is suitable for the behavior recognition of students in the classroom.

**4.2.2 Module 2**

The main function of Module 2 is to complete the automatic evaluation of the teacher's teaching ability through the computer. First, the speech recognition model is used to convert the teacher's in-class speech data into text data, then the text data is input into the large language model to complete the evaluation, and finally the evaluation report is output.

This paper selects the SenseVoiceSmall model as the speech recognition model. This model is a Chinese-optimized speech recognition model based on a non-auto-regressive framework, which supports the conversion of audio data of multiple languages such as Chinese, English, Cantonese, and Japanese into text data, and has the characteristics of short conversion time and low error rate. Compared with the international general speech recognition model Whisper-large, it has higher recognition accuracy and an operating speed about 23 times faster.

After converting the teacher's teaching audio data into text data, the relevant corpus analysis is completed based on the large language model. This paper selects the DeepSeek-R1:70b version as the analysis model for text data. Through prompt instructions, the local large language model is allowed to first summarize the teacher's teaching corpus, and then complete the evaluation and analysis of the corpus data from three dimensions: classroom ideology and politics, teaching content, and teaching methods.

**4.2.3 Module 3**

This module is the classroom data mapping analysis module, which uses the students' in-class performance (head-up rate) as the judgment index to carry out positive or negative annotation on the teaching corpus data, stores the annotation results and corpus data in the teaching database, and generates the corresponding teaching optimization suggestion report. The execution process is shown in Figure 5.

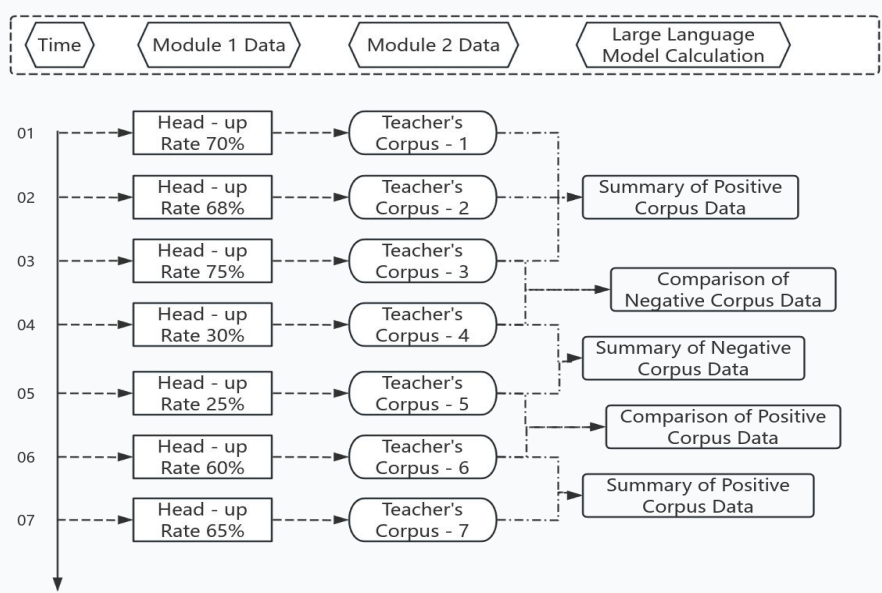

Fig. 5 Teaching Analysis Module Flowchart.

First, the statistical results of students' classroom behavior recognition in Module 1 are extracted, and the head-up rate of each recognition result is calculated according to the number of students looking up and bowing their heads in each recognition result. The average value of all recognition rates within one minute is taken to obtain the head-up rate of students within any one minute in the classroom. The head-up rate data is arranged in chronological order and drawn into a chart to show the changing trend of students' group behavior during the class.

Then, according to the changing trend of the head-up rate per minute, the time axis of classroom teaching is divided into intervals and special points are marked. The intervals are divided into: high head-up rate stage, medium head-up rate stage, and low head-up rate stage, and the special points are marked as the time points where the head-up rate significantly increases or decreases. The teacher's corpus data in Module 2 is corresponding to the time interval, marked as high/medium/low head-up rate corpus data, and stored in the database. According to the special time points, the corpus data before and after the time points where the head-up rate significantly increases/decreases is extracted, marked as positive/negative contrast corpus data, and stored in the database.

Finally, based on the DeepSeekR1 local large language model, the high/medium/low head-up rate corpus data and positive/negative contrast corpus data in the database are analyzed and summarized. The high head-up rate corpus data indicates that the teaching content and expression form have high attractiveness and adaptability to students, and vice versa. The positive contrast corpus data is the contrast between the high head-up rate corpus data and the low head-up rate corpus data. Through the contrast of teaching content and expression form in the corpus data with different head-up rates, a more efficient information expression form and more appropriate teaching content can be analyzed. Through the analysis and comparison of the above different labeled data, the influence of different teaching contents, expression forms, and other factors on the teaching effect is obtained, and the optimization suggestion report and data storage are generated.

## 5. Classroom Evaluation Practice

This paper selects the Advertising History course of a certain university as the analysis material to carry out the practice test of the classroom comprehensive evaluation system in this study. First, in Module 1, the image recognition is carried out on the video, and the change trend of the number of students looking up and bowing their heads in the whole class is counted. The recognition results are shown in Figure 6.

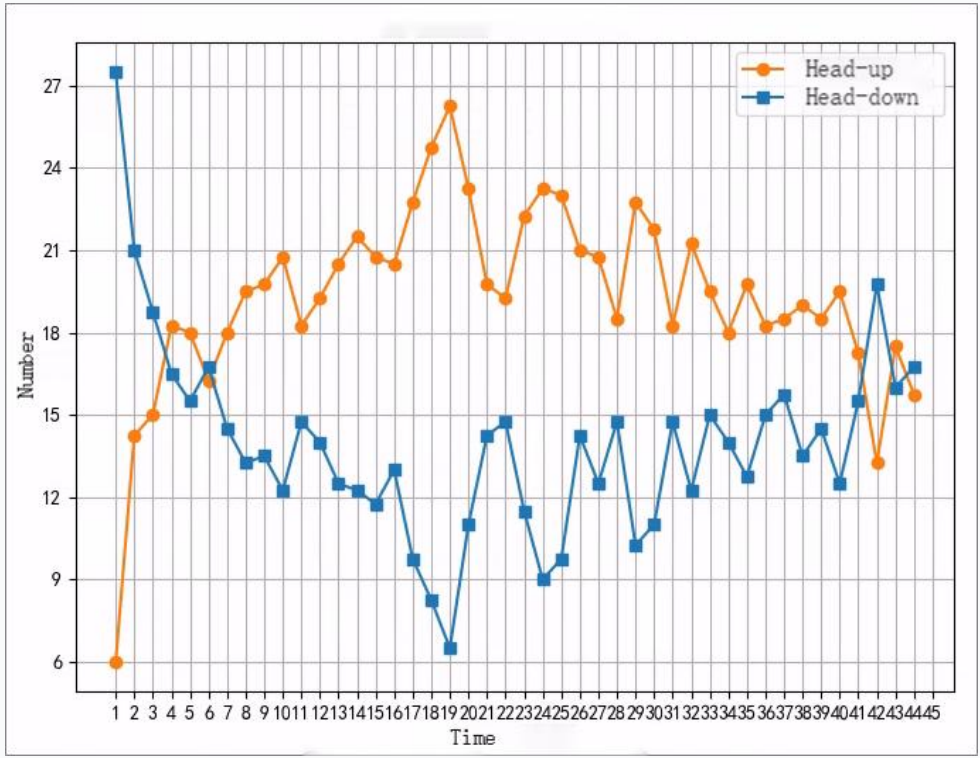

Fig. 6 Trend Chart of Students' Classroom Behaviors.

The statistical results show that from the 11th to the 19th minute of the class, the head-up rate of the student group showed an obvious upward trend; from the 19th to the 22nd minute, the head-up rate of the student group showed a downward trend; from the 22nd to the 40th minute, the head-up rate of the student group was in a stable trend. Based on the statistical analysis of the number of students looking up and bowing their heads, the number of participants in this class was 35. The average number of students looking up per minute in the whole class was 19.39, the average number of students bowing their heads was 13.87, and the average ratio of looking up to bowing their heads was 1.40, that is, the average number of students looking up in the whole class was about 1.4 times that of bowing their heads. The evaluation of the students' learning performance in the whole class is good.

Then, the speech recognition is carried out on the original video to obtain the teacher's teaching corpus text per minute, and then the corpus text is combined with the selected evaluation indicators and input into the local language model to generate the classroom evaluation report, as shown in Table 1.

Table. 1 Class evaluation report.

| Evaluation Dimension | Conclusion | Analysis |
| --- | --- | --- |
| 1. Summarize the main teaching content of the class. | This class mainly focuses on the development history of the advertising industry, from the Qing Dynasty to the modern era. It highlights the development of China's advertising industry in different historical periods | The course content covers the history of advertising development from the Qing Dynasty to the period of the War of Resistance against Japan, and especially emphasizes Soong May |
| | and its relationship with politics, economy, and culture. Specifically including: ① The transformation from a feudal society to a modern commodity economy; ② The emergence of commercial entities in cities such as Shanghai during the Republic of China period; ③ Soong May - ling's speech in the US Congress, demonstrating the combination of advertising and political propaganda. | - ling's role on the international stage. |
| 2. Is there integration of ideological and political education in the class? | Yes. The teacher integrated ideological and political elements during the lecture, such as the cooperation between the Kuomintang and the Communist Party in the War of Resistance against Japan, and the historical roles of Chiang Kai - shek and Soong May - ling, conveying the spirit of patriotism and a sense of historical responsibility. | The lecturer mentioned the cooperation between the Kuomintang and the Communist Party in the War of Resistance against Japan, and the roles of Chiang Kai - shek and Soong May - ling. These contents are related to political events in modern Chinese history and may be integrating patriotism education and a sense of historical responsibility. |
| 3. Is the teaching logic of the class smooth and rigorous? | The teaching logic is not rigorous enough. The teacher made quite a few jumps during the lecture. For example, suddenly switching from the development of the advertising industry to Soong May - ling's speech without proper transition, which may make students confused. | The teacher's topic transition is rather abrupt. For instance, switching directly from commodity economy to Soong May - ling's speech, and the logical connection is not tight enough. |
| 4. Does the teaching process of the class include the combination of theory and practice in the explanation? | Yes. The teacher combines advertising theory with historical reality through commodity economy, urbanization, the case of Soong May - ling, etc., to help students understand the role of advertising in different historical stages. | The teacher used specific historical events and character cases, such as Soong May - ling's speech, to illustrate the combination of advertising and political propaganda, reflecting the combination of theory and practice. |
| 5. Is the subject characteristic obvious? | Obvious. The course content involves the origin and development of the advertising industry, which is closely combined with modern Chinese history. It has strong interdisciplinary | The course not only covers advertising theory but also deeply explores its role in modern Chinese history, intersecting with other disciplines such as history |

characteristics, highlighting the connection between advertising science and history, politics. and political science, making the course more unique and in-depth.

Finally, according to the trend statistical results of students' classroom behaviors, the corpus data in specific time intervals is selected and sent to the local language model to generate the optimization suggestion report, as shown in Table 2.

Table. 2 Teaching optimization recommendation report.

| Student Behavior | Teaching Content and Expression Forms | Analysis |
|---|---|---|
| 11 - 19 minutes: Student head-up rate increases | Content: During this period, the teacher talked about the "Golden Age" in the history of advertising, specifically covering the development from the early 20th century to the 1940s. It included classic cases such as Coca-Cola and Ford Motor, as well as advertising techniques and technological innovations.<br><br>Expression forms: 1. The teacher used the success stories of specific brands like Coca-Cola and Ford Motor, making it easy for students to resonate. 2. Told how advertising affected people's emotions and lives, stimulating students' interest in the connection between history and reality. 3. Probably used pictures or videos to display classic advertisements, helping students understand better. | 1. Specific cases enhanced the attractiveness of the content, making it easier for students to get involved.<br>2. Emotional and story elements increased participation, enlivening the classroom atmosphere.<br>3. Visual aids made information processing more effective, keeping students' attention focused. |
| 19 - 22 minutes: Head-up rate decreases | Content: This stage shifted to the development of advertising from the 1950s to the 1970s, including the rise of TV advertising and the application of new technologies. Discussed consumerism and social responses.<br><br>Expression forms: 1. The content was rather theoretical, involving concepts like consumerism, lacking the story-telling nature of the previous stage. 2. Shifted to analytical content, reducing students' emotional investment. | 1. The content changed from vivid cases to abstract theories, causing some students to lose interest.<br>2. The lack of emotional connection and specific examples led to a decline in participation.<br>3. Compared with the high interactivity of the previous stage, this part of the content was rather monotonous, and students' attention was scattered. |
| 22 - 40 minutes: | Content: During this period, it mainly | 1. The content was biased |
| Student head-up rate stabilizes at 55% | talked about the history of modern Chinese advertising, from the Xinhai Revolution to the War of Resistance against Japan. Discussed the commercial development of cities like Shanghai and Wuhan and the influence of celebrities such as Soong May-ling. | towards historical narration, and the connection with the practical application of advertising history was not close enough.<br>2. The case analysis was rather superficial and failed to deeply explore its impact on modern advertising.<br>3. The lack of interactive links and emotional elements made some students feel that the content was boring. |
| Summary | Expression forms: 1. Involved political changes and the transformation of economic subjects, and the content was rather academic. 2. Such as Soong May-ling's experience of giving a speech in the United States, but there was a lack of in-depth discussion. 3. Mainly based on telling, with limited student participation.<br>Through specific cases, emotional connections, and visual aids, the teacher increased students' participation from 11-19 minutes. Then it shifted to theoretical analysis, and the lack of story-telling led to a decline in interest. From 22-40 minutes, although the historical background and character cases were attractive, the rather academic way of telling limited the further improvement of participation. It is recommended to add interactive links, such as discussions or practical activities, to maintain the overall enthusiasm of students. | |

After the above three links, the system generates a teaching effect and teaching ability evaluation report from the perspective of students' behavior feedback and the teacher's own ability for the classroom teaching process. Finally, by mapping the relationship between students' behavior data and the teacher's corpus data, an optimization suggestion report is generated, forming a closed-loop system for comprehensive evaluation of teaching activities and guiding teaching activities based on data analysis. The test material this time was 45 minutes long. In the teaching effect evaluation stage based on image recognition, the computer took 19 seconds; in the speech recognition stage, the computer took 20 seconds; in the local language model execution text summarization stage, the computer took 58 seconds. Compared with the existing classroom supervision full-class listening evaluation method, this study has significant advantages in evaluation efficiency. In the future, through multi-threaded program parallel computing and increasing the number of computers, the efficiency can be further improved to complete popularization and promotion.

## 6. Conclusion

The development trend of panoramic intelligence in education and teaching requires the digital transformation of teaching evaluation. With the help of big data and artificial intelligence technologies, the teaching process should be analyzed from a more comprehensive, in-depth, and systematic perspective to guide educational practice to continuously improve in a more scientific way. Based on the purpose that teaching evaluation should guide teaching reform, this paper designs a closed-loop evaluation model from evaluation results to guiding teaching, and completes the development of a comprehensive classroom evaluation system by combining multiple artificial intelligence technologies. The system can automatically complete the multi-dimensional evaluation of teaching ability and teaching effect according to the teacher's corpus and students' behavior performance of each class based on the teaching objectives, and generate corresponding optimization suggestions. Through the teaching evaluation results and optimization suggestions of each class, the process visualization of promoting the improvement of classroom teaching quality is realized, providing support for the internal drive of teacher growth, and providing practical tools for the improvement of the overall teaching level in China. While completing the teaching evaluation, the teaching data collection function of this system can also provide corresponding data support for the analysis of teaching mechanisms, the exploration of value-added evaluation, and the effectiveness of various educational policies, promoting the comprehensive collaborative development of various fields of education and teaching.